# Awaking of ferromagnetism in GaMnN through control of Mn valence

Saki Sonoda[a)]

*Department of Electronics and Information Science, Kyoto Institute of Technology*

*1-1 Goshokaido, Matsugasaki, Sakyo, Kyoto 606-8585, Japan*

Isao Tanaka, Fumiyasu Oba, Hidekazu Ikeno, Hiroyuki Hayashi

*Department of Materials Science and Engineering, Kyoto University*

*Yoshida, Sakyo, Kyoto 606-8501, Japan*

Tomoyuki Yamamoto

*Department of Material Science and Engineering, Waseda University*

*3-4-1 Okubo, Shinjuku, Tokyo 169-8555, Japan*

Yoshihiko Yuba, Yoichi Akasaka

*Department of System Innovation, Osaka University*

*1-3 Machikaneyama, Toyonaka,Osaka 560-8531 Japan*

Ken-ichi Yoshida, Masahiko Aoki, Masatoshi Asari,

*Ion Enginieering Research Institute Corporation*

*2-8-1 Tsuda, Yamate, Hirakata, Osaka 573-0128, Japan*

Koichi Kindo

*The Institute for Solid State Physics, University of Tokyo*

*5-1-5 Kashiwanoha, Kashiwa, Chiba 277-8581, Japan*

Hidenobu Hori

*School of Material Science, Japan Advanced Institute of Science and Technology*

*1-1 Asahidai, Nomi, Ishikawa 923-1292 Japan*

Room temperature ferromagnetism of GaMnN thin film is awaked by a mild hydrogenation treatment of a sample synthesized by molecular beam epitaxy. Local environment of Mn atoms is monitored by Mn-$L_{2,3}$ near edge x-ray absorption fine structure technique. Doped Mn ions are present at substitutional sites of Ga both before and after the hydrogenation. No secondary phase can be detected. Major valency of Mn changes from +3 to +2 by the hydrogenation. The present result supports the model that the ferromagnetism occurs when $Mn^{2+}$ and $Mn^{3+}$ are coexistent and holes in the mid-gap Mn band mediate the magnetic coupling.

[a)] To whom correspondence should be addressed. E-mail sonoda@kit.ac.jp

In order to realize practical spin-electronics devices, it is essential to have a diluted magnetic semiconductor (DMS) working at room temperature (RT). Recently many reports claimed their success of RT ferromagnetism in wide-gap matters.[1] However, mechanism behind the ferromagnetism is still controversial. Clear guiding principles to make the room temperature ferromagnetism have not been established.

Recently the present authors examined Mn doped GaN (GaMnN) showing room temperature ferromagnetism and $p$-type conductivity.[2] The valence of Mn was monitored by the Mn-$L_{2,3}$ near edge x-ray absorption fine structure (NEXAFS). Coexistence of $Mn^{2+}$ (major) and $Mn^{3+}$ (minor) was demonstrated in the ferromagnetic GaMnN films. The ferromagnetism of the film vanishes at low temperatures where its carrier concentration decreases significantly. It was also noted that the ferromagnetism does not show up when $Mn^{3+}$ is predominant.

The present experiments are designed in order to provide further proof of this model. On the basis of the model, magnetism can be manipulated by the redox control of Mn ions. Paramagnetic or very weak ferromagnetic GaMnN samples with predominantly $Mn^{3+}$ can be converted into ferromagnets when the valence is controlled to make $Mn^{2+}$ and $Mn^{3+}$ coexistent. The most popular approach to change the valence may be co-doping of other impurities during the synthesis of thin film samples. However, optimum process conditions are often greatly affected by the presence of the co-dopants. Unintentional side-effects may cover up the real effects unless the optimum conditions are chosen. In the present study, we have selected a mild hydrogenation treatment in order to introduce hydrogen atoms into the GaMnN thin film. Thin films before and after the hydrogenation process were subjected to Mn $L_{2,3}$ NEXAFS and secondary ion mass spectroscopy (SIMS) analyses to examine the local environment and valence of Mn together with the content of H. Magnetization was measured by superconducting quantum interference device (SQUID).

GaMnN thin films, 300 nm in thickness, with Mn concentration ranging from 6 to 8cation% were prepared by molecular beam epitaxy (MBE) using $NH_3$ as nitrogen source as described elsewhere.[3,4] In order to obtain films that show paramagnetic or weak ferromagnetic behaviour at RT, we applied relatively high substrate temperatures during the growth. For example, under the fixed Ga and Mn fluxes, the film as grown at 720°C (Mn:6.8%) shows ferromagnetism at RT,[2] while one grown at 820°C shows paramagnetism at RT in spite of a similar Mn concentration (6.0%).

The films were then exposed to atomic hydrogen (H*). Figure 1 shows schematic of the atomic H* exposure system employed in the present study.[5] H* was generated by cracking of $H_2$ with a hot tungsten (W) wire. No electric/magnetic fields were applied. A hydrogenation cycle was composed of 60 min H*-exposure with 30 min idle time. The temperature of the film during the hydrogenation process was carefully measured not to exceed $100\,^{\circ}C$. Since the grown films have cation (Ga) polarity, the film surfaces were thought not to be decomposed in the temperature range.[6] In fact, even after 10-cycles exposure, no change of the GaMnN film thickness can be detected. Concentration profile of hydrogen as monitored by SIMS is shown in Fig. 2. The H concentration in the H*-exposed sample is largest at the surface, which is the order of $1 \times 10^{22}$ cm$^{-3}$ at the depth of 6.8nm, This corresponds to 30% of the number of cations in GaMnN. The H concentration is more than a factor 20 greater than the surface of the original sample. It decreases with depth and almost disappears at around 100nm, where the H concentration is the order of $1 \times 10^{19}$ cm$^{-3}$.[7] The hydrogen from the external source does not diffuse into the deeper region.

Figure 3 shows Mn-$L_{2,3}$ NEXAFS from two samples measured by the total electron yield method at BL11A in KEK-PF, Tsukuba, Japan. The spectra provide information of specimen surface less than a few nm in depth. The fine structures of the spectra are interpreted on the basis of first-principles multi-electron calculations as reported previously.[2] The results show that the doped Mn ions are present at substitutional sites of Ga both before and after the hydrogenation. Contribution from other phases cannot be detected, which is consisted with XRD analysis performed for both specimens. The change in the spectrum implies that the valence of Mn is changed from 3+ to 2+ by the hydrogenation in order to compensate the charge of $H^+$ at the interstitial sites.

Magnetic properties of the films measured by SQUID at RT before and after the hydrogenation are summarized in Fig.4 (a) and (b). Figure 4(a) shows raw magnetization data of a sample which exhibits almost paramagnetism before the hydrogenation. Linear background magnetization with negative gradient can be ascribed mainly to diamagnetism of the sapphire substrate, which is predominant in the as-grown sample. After the hydrogenation, a hysteresis loop appears, implying ferromagnetism is induced. With the increase of the number of hydrogenation cycles, the hysteresis loop becomes larger.

Figure 4(b) shows the magnetization behavior of another sample which exhibits weak ferromagnetism in the as grown-state. In this case, the linear background magnetization has been subtracted in order to extract the ferromagnetic contribution from the raw data. The ferromagnetism is found to be reinforced with the hydrogenation cycle also in this sample.

We hereby demonstrate that the magnitude of the ferromagnetism in the GaMnN can be manipulated by the control of the valence of magnetic impurities. This may be possible not only by the hydrogenation but also by co-doping of secondary impurities or simply by the change in process conditions to modify the defect chemistry.

This work was supported by three projects by Japanese Ministry of Education, Culture, Sports, Science and Technology (MEXT). They are the computational materials science unit in Kyoto University, the Grant-in-Aid for Scientific Research

Figure caption

Fig. 1　H* exposure system. $H_2$ gas was introduced into the chamber through a variable leak valve, and H* was generated by cracking $H_2$ with a hot tungsten (W) wire.

Fig. 2　SIMS profile of hydrogen for GaMnN (Mn: 6.0 %) thin film before and after 10 cycles hydrogenation.

Fig. 3　Mn $L_{2,3}$-NEXAFS of GaMnN (Mn: 6.0 %) thin film.

Fig. 4　(a) Raw data of magnetization vs. magnetic field curve of GaMnN (Mn: 6.0%). (b) Magnetization vs. magnetic field curve of GaMnN (Mn: 7.7%) after subtraction of the linear background.

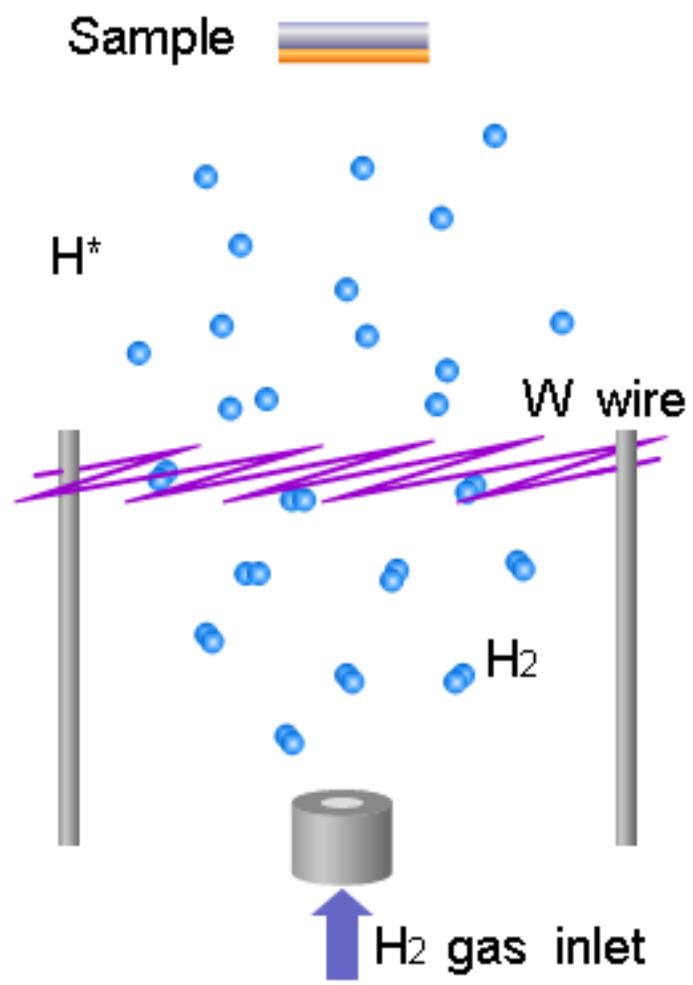

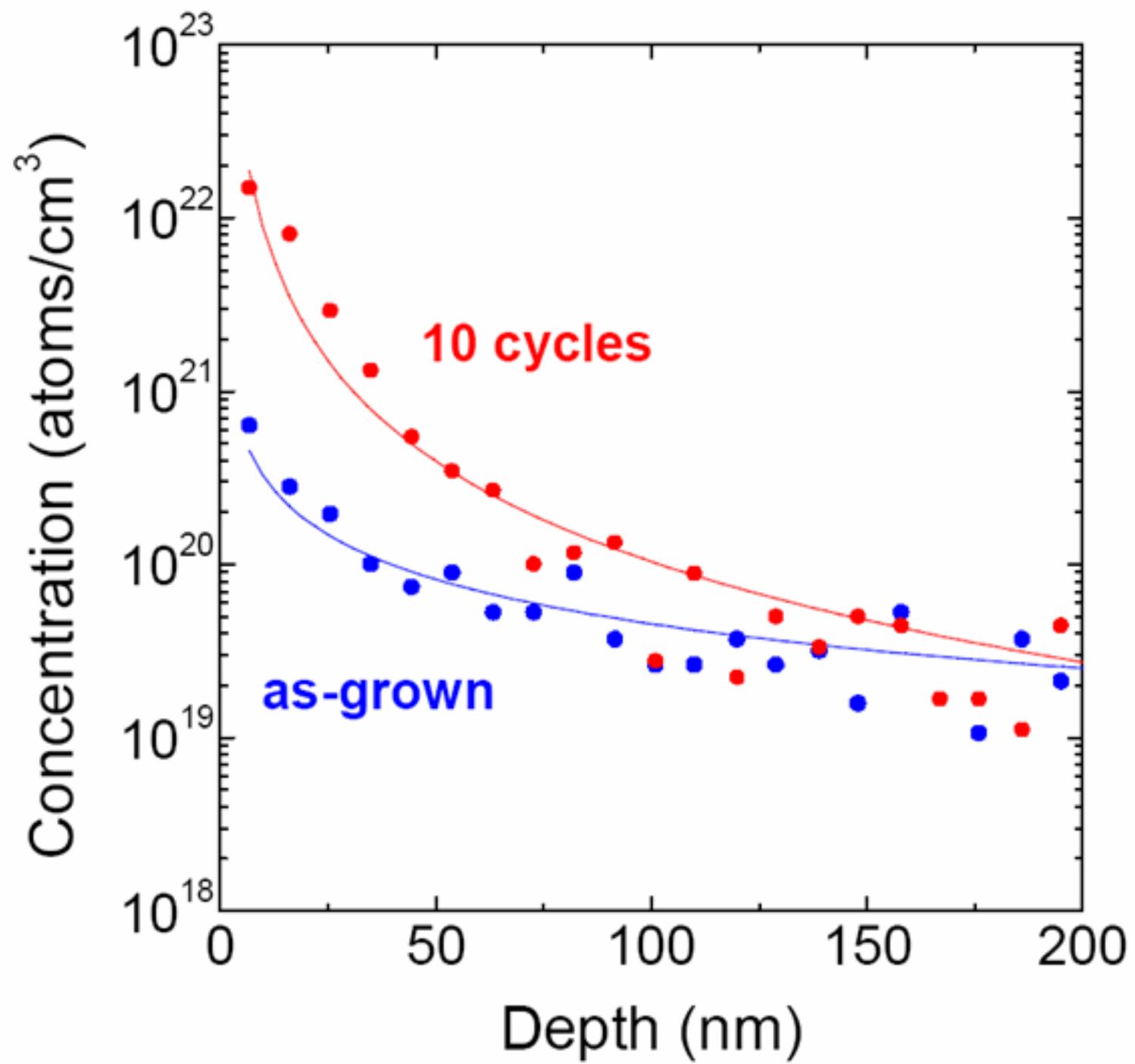

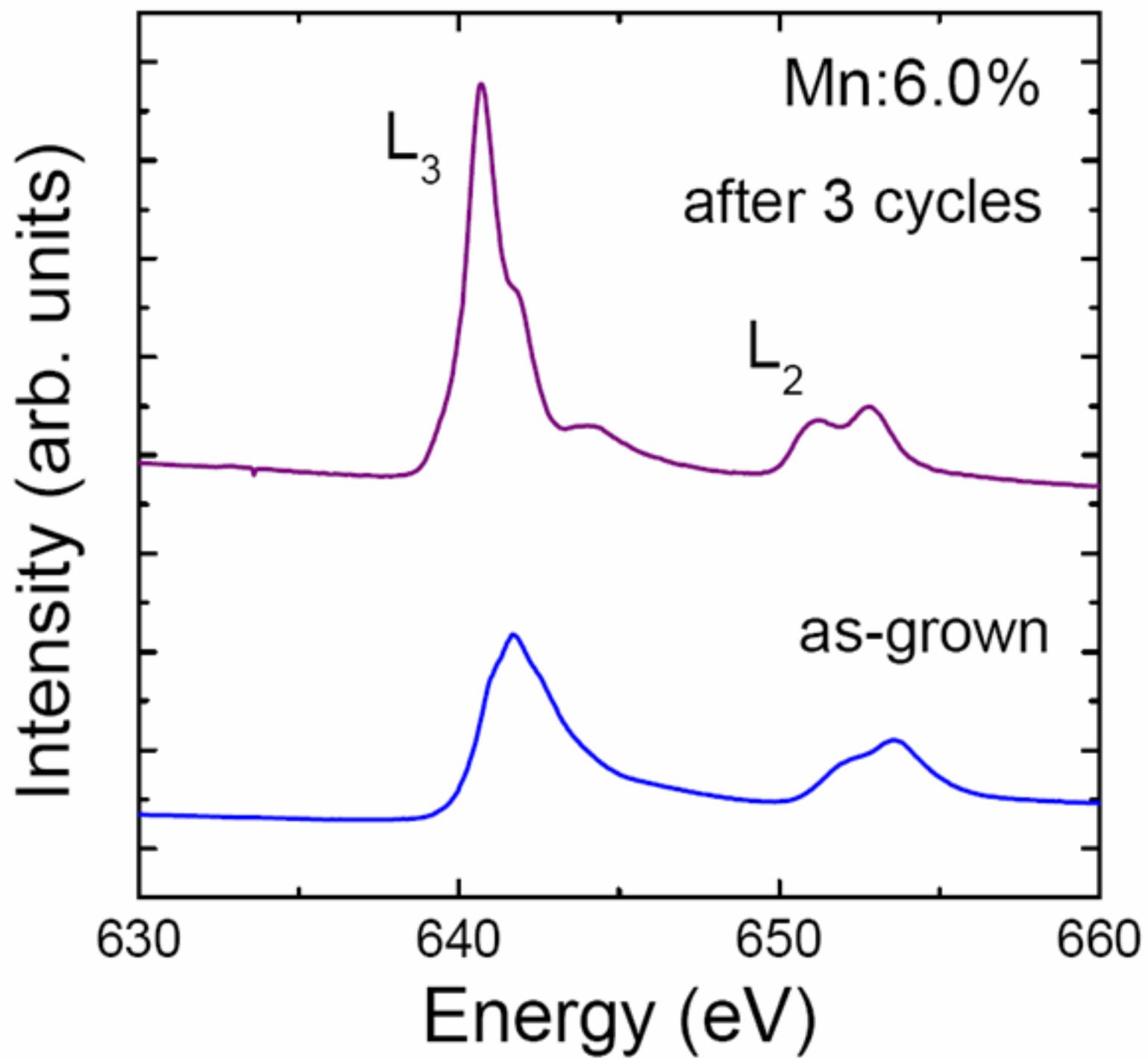

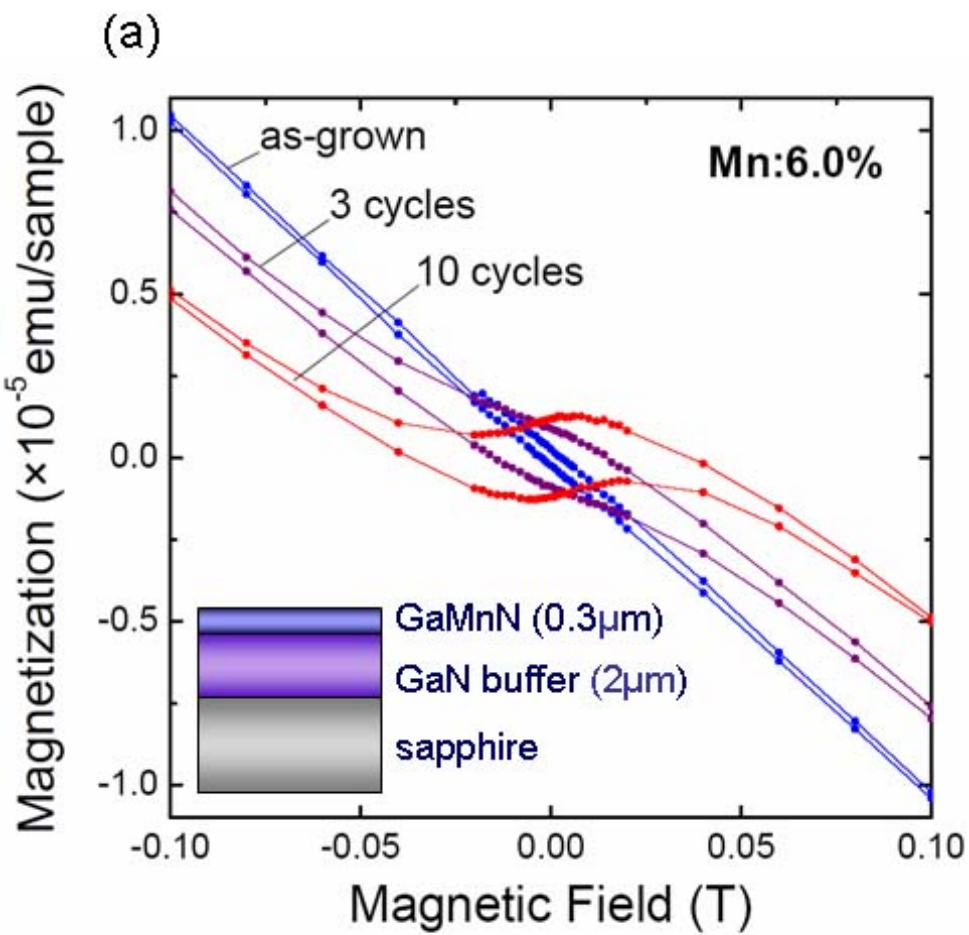

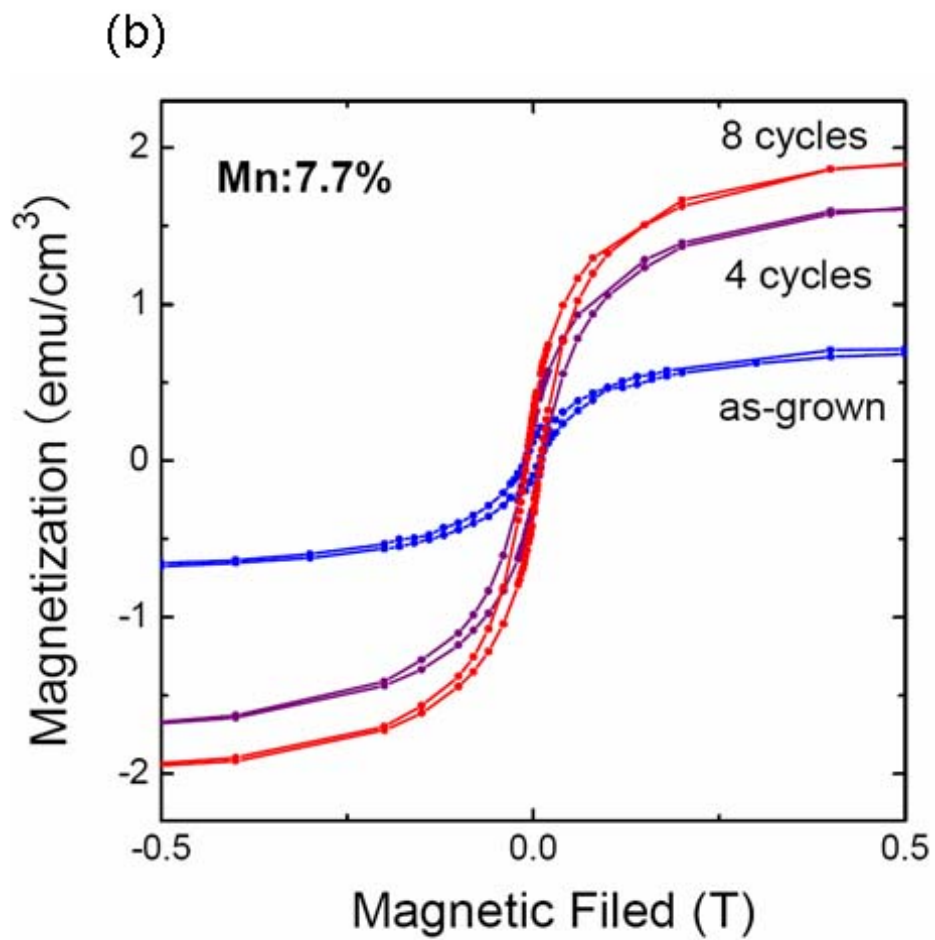